\documentclass{caps}
\usepackage{peter}
\usepackage{graphicx}
\usepackage{epsf}
\def\note #1]{{\bf #1]}}
\def\note #1]{{\bf #1]}}

\begin{document}

\author{David Branch\\ Department of Physics and Astronomy, University
of Oklahoma, Norman, OK 73019, USA}

\chapter{Type Ia supernovae: spectroscopic surprises}

{\it Recent observations have extended the range of diversity among
 spectra of Type~Ia supernovae. I briefly discuss SN~Ia explosion
 models in the spectroscopic context, the observed diversity, and some
 recent results obtained with the {\bf Synow} code for one normal and
 two peculiar SNe~Ia.  Relating the observational manifestations of
 diversity to their physical causes is looking like an ever more
 challenging problem.}

\section{Introduction}

``Surprises'' refers not only to some recent developments in Type~Ia
supernova (SN~Ia) spectroscopy that will be discussed below, but also
to additional recent discoveries that I will be able only to mention,
such as the polarization signal in SN~2001el (Wang et~al. 2003; see
also the chapter by Wang); the unusual properties of SN~2001ay (see
the chapter by Howell); and the circumstellar H$\alpha$ emission of
SN~2002ic (Hamuy et~al. 2003; see also the chapter by Hamuy).  The
scope of this chapter is restricted to photospheric--phase optical
spectra.  For recent results on infrared spectra see, e.g., Marion
et~al. (2003).

Some background, including mention of the various kinds of SN~Ia
explosion models in the spectroscopic context, is in \S1.2.  An
overview and update of the SN~Ia spectroscopic diversity is in \S1.3.
Some recent results from direct analysis of the spectra of three
events (the normal SN~1998aq and the peculiar SNe~2000cx and 2002cx),
obtained with the parameterized, resonance scattering code {\bf
Synow}, are discussed in \S1.4.  The final section (\S1.5) contains
more questions than conclusions.

\section{Background}

Around the time of maximum light the optical spectrum of a normal
SN~Ia consists of a thermal continuum with superimposed features due
to lines of ions such as Si~II, S~II, Ca~II, and O~I. Ubiquitous
blends of Fe~II develop shortly after maximum.  The features are
formed by resonance scattering of the photospheric continuum and have
P~Cygni--type profiles characteristic of expanding atmospheres.
Emission components peak at the rest wavelength and absorption
components are blueshifted according to the velocity of the matter at
the photosphere, ordinarily $\sim10,000$ km~s$^{-1}$.

SNe~Ia are thought to be thermonuclear disruptions of accreting or
merging carbon--oxygen white dwarfs.  The classic SN~Ia explosion
model is model~W7 of Nomoto, Thielemann, \& Yokoi (1984), a 1D model
that was constructed by parameterizing the speed of the nuclear
burning front.  Because the speed remained subsonic, W7 is known as a
deflagration model (but see below).  The composition structure of W7
is radially stratified, with a low-velocity ($<10,000$ km~s$^{-1}$)
$\sim$1~M$_\odot$ core of iron-peak elements (initially mostly
radioactive $^{56}$Ni), surrounded by $\sim$0.3~M$_\odot$ of
intermediate-mass elements such as silicon, sulfur, and oxygen
expanding at $\sim10,000$ to $\sim15,000$ km~s$^{-1}$, capped by
$\sim$0.1~M$_\odot$ of unburned carbon and oxygen moving at
$\sim15,000$ to $\sim22,000$ km~s$^{-1}$.  Synthetic-spectrum
calculations of various levels of complexity (Branch et~al. 1985;
Wheeler \& Harkness 1990; Nugent et~al. 1995; Salvo et~al. 2001; Lentz
et~al. 2001a) have shown that the spectra of W7, with or without
mixing of the layers above $\sim8000$ km~s$^{-1}$, generally resemble
the observed spectra of normal SNe~Ia.  Recent detailed calculations
with the {\bf Phoenix} code indicate that spectra of model~W7 begin to
differ from observed spectra after maximum light (Lentz et~al. 2001a),
and that they may fail to quantitatively reproduce the dependence of
the $R(Si~II)$ parameter on temperature (S.~Bongard et~al., in
preparation).  Nevertheless, there seems to be something right about
the basic composition structure of W7: an iron--peak core surrounded
by lighter elements.

In 1D delayed--detonation models (Khokhlov 1991) the subsonic
deflagration makes a transition to a supersonic detonation when the
burning front reaches a transition density that is treated as a free
parameter.  Composition structures of 1D delayed--detonations are not
extremely different from that of model~W7.  The main differences are
in the outer, high--velocity layers, because in delayed--detonations
the burning front extends farther into the outer layers of the white
dwarf and leaves less unburned carbon than W7.  [A nice comparison of
the composition structures of W7 and a 1D delayed--detonation can be
found in Sorokina et~al. (2000).]  H\"oflich and colleagues (e.g.,
H\"oflich 1995; H\"oflich, Wheeler, \& Thielemann 1998; Wheeler
et~al. 1998; see also the chapter by H\"oflich) have maintained that
spectra of delayed--detonation models agree well with the observed
spectra of normal SNe~Ia.

Only recently have 3D deflagrations been calculated (Khokhlov 2000;
Gamezo et~al. 2003; see also the chapter by Gamezo, Khokhlov, \& Oran
and references therein).  The composition structure of the 3D
deflagration presented by Khokhlov and Gamezo et~al. is quite unlike
that of a 1D deflagration.  The 3D model contains clumps of iron--peak
elements surrounded by shells of intermediate--mass elements, all
embedded in a substrate of unburned carbon and oxygen.  The
angle--averaged composition is {\sl not} strongly radially stratified;
in fact, unburned carbon is present all the way to the center.
Although the model has not yet been evolved to homologous expansion,
it is unlikely that it will be compatible with the observed spectra
of normal SNe~Ia.  The presence of unburned carbon at low velocity may
not be a problem during the first months after explosion (Baron
et~al. 2003; see also the chapter by Lentz), but a strong argument can
be made for a smooth distribution of silicon in normal SNe~Ia in order
to consistently produce the deep 6100~\AA\ absorption (Thomas
et~al. 2002; see also the chapter by Thomas).  The 3D model also would
produce [C~II] and [O~I] lines in the nebular phase (C.~Kozma et~al.,
in preparation), which are not observed.  However, some future 3D
deflagration models may be different: much may depend, for example, on
whether the deflagration originates at a single point or practically
simultaneously at several or even a multitude of points.  

In the chapter by Gamezo et~al. a 3D {\sl delayed--detonation} model
is discussed, in which the composition structure {\sl is} radially
stratified, rather like 1D explosion models.  If the composition
structures of future 3D deflagrations continue to be very different
from that of model~W7, perhaps W7 should be referred to as a
parameterized 1D model, rather than as a deflagration, because its
composition structure may be more like 3D delayed--detonations.

The effects on explosion models of rapid rotation of the white--dwarf
progenitors (Uenishi et~al. 2003; see also the chapter by Yoon and
Langer) has not yet begun to be taken into account.

\section{Spectroscopic diversity}

Nugent et~al. (1995) showed that retaining the composition structure
of model~W7 and varying the temperature produced a sequence of
synthetic spectra that resembled a sequence of observed SN~Ia spectra
ranging from the peculiar powerful SN~1991T, with its high--excitation
Fe~III features, through the normals, to the peculiar weak SN~1991bg,
with its low--excitation Ti~II features.  Quantitatively, the sequence
was arranged according to increasing values of the parameter
$R(Si~II)$, the ratio of the depth of an absorption feature near
5700~\AA\ to the depth of the deeper absorption near 6100~\AA\ that is
primarily due to Si~II $\lambda$6355. [The major contributors to the
5700~\AA\ aborption remain uncertain (Garnavich et~al. 2001,
S.~Bongard et~al., in preparation)]. As shown by Nugent et~al. (1995),
Garnavich et~al. (200?), and Benetti et~al. (2003), the $R(Si~II)$
parameter correlates with the light--curve decline--rate parameter
$\Delta m_{15}$ of Phillips (1993).

But it has become well known that not all SNe~Ia can be forced into a
single spectroscopic sequence.  Events such as SN~1984A have the same
spectral features as normal SNe~Ia but higher expansion velocities at
a given epoch.  I will refer to the most extreme of these as HPV (high
photospheric velocity) SNe~Ia.  Lentz et~al. (2001b) found that
certain delayed--detonation models, with their high densities in the
high--velocity ($\sim20,000$ km~s$^{-1}$) layers, were able to account
well for the spectra of SN~1984A.  The HPV SNe~Ia have high values of
the $V_{10}(Si~II)$ parameter --- the velocity corresponding to the
blueshift of the Si~II $\lambda$6355 absorption at 10~days after
maximum light (Branch \& van den Bergh 1993).  Hatano et~al. (2000)
presented a plot of $R(Si~II)$ against $V_{10}(Si~II)$ to illustrate
the extent to which some events having similar values of $R(Si~II)$
have different values of $V_{10}(Si~II)$.  Hatano et~al. suggested
that it may be useful to think of $R(Si~II)$ (like $\Delta m_{15}$) as
a measure of the ejected nickel mass and $V_{10}(Si~II)$ as a measure
of the density in the high--velocity layers --- whether the high
density actually is produced by a delayed detonation or not.
Similarly, one can plot $\Delta m_{15}$ against $V_{10}(Si~II)$, as in
Fig.~1.  Plotting $\Delta m_{15}$ allows some interesting events for
which $R(Si~II)$ is unavailable to be included.  The dashed line
serves to represent the Nugent ``temperature'' sequence connecting the
peculiar SN~1991bg to the peculiar SN~1991T via the normal SN~1994D (a
different normal event, e.g., SNe~1989B, 1996X, or SN~1998bu, could
just as well have been chosen).  SN~2001ay, an HPV event with a very
slow postmaximum light--curve decline (see the chapter by Howell) and
SN~2002cx (Li et~al. 2003) have extended the observed range of $\Delta
m_{15}$ and $V_{10}(Si~II)$, respectively, and the value of
$V_{10}(Si~II)$ plotted for SN~2002cx is an upper limit because the
Si~II absorption was seen only near maximum light.


\begin{figure}
\includegraphics[width=8cm, angle=270]{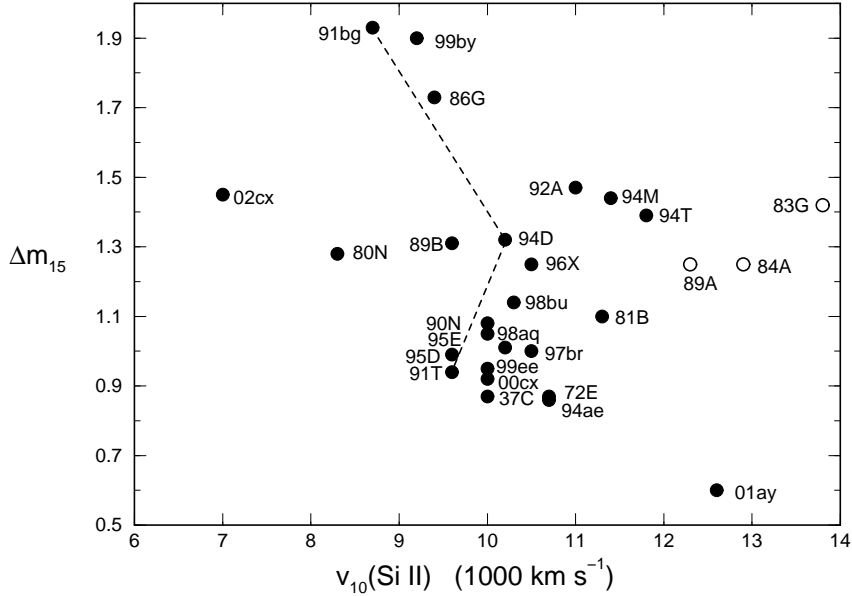}
\caption{The photometric $\Delta m_{15}$ parameter is plotted against
  the spectroscopic $V_{10}(Si~II)$ parameter.  Open circles mean that
  the value of $\Delta m_{15}$ has been estimated from the observed
  value of the $R(Si~II)$ parameter. The dashed line is intended to
  represent the temperature sequence of Nugent et~al. (1995).}
\end{figure}

\section{Recent results with {\bf Synow}}

The spectroscopically normal SN~1998aq was thoroughly observed by the
CfA group.  In Fig.~1 it falls in a region heavily populated by
normals.  One result of a recent direct analysis of the
photospheric--phase spectra with the {\bf Synow} code (Branch
et~al. 2003) was good evidence for the presence of C~II lines
indicating the presence of carbon at velocities at least as low as
11,000 km~s$^{-1}$.  Mazzali (2000) identified the same lines, but at
higher velocity, $\sim16,000$ km~s$^{-1}$, in the normal SN~1990N.
Carbon as slow as 11,000 km~s$^{-1}$ would be inconsistent with
published 1D delayed--detonation models.  Most of the other spectral
features in the early spectra were securely identified, making
SN~1998aq a useful benchmark for comparison with other SNe~Ia.

SN~2000cx was extensively observed by Li et~al. (2002), who referred
to it as uniquely peculiar.  In Figure~1 it falls among the normals,
but the unusual properties discovered by Li et~al. included a lopsided
$B$--band light curve --- quick to rise but slow to fall --- and a
strange $B - V$ color evolution --- from redder than normal before
maximum to bluer than normal after maximum.  The many spectral
peculiarities included Ca~II infrared--triplet absorptions forming not
only at the photospheric velocity of $\sim10,000$ km~s$^{-1}$ but also
at much higher velocity, greater than 20,000 km~s$^{-1}$.  Thomas et
al. (2003) investigated the high--velocity Ca~II features with a
parameterized 3D spectrum--synthesis code and concluded that they
probably formed in a nonspherical high--velocity structure that
partially covered the photosphere.  Kasen et~al. (2003) already had
reached a similar conclusion for the high--velocity Ca~II features
discovered by Wang et~al. (2003) in SN~2001el, for which polarization
measurements clearly indicated a strong departure from spherical
symmetry in the high--velocity matter.  I will refer to features like
these as DHVFs (detached high velocity features).  Direct analysis of
the photospheric--phase spectra of SN~2000cx with {\bf Synow}
(D.~Branch et~al., in preparation) reveals that in the blue, the
spectra were composite, containing some features forming just above
the photospheric velocity, but also containing additional DHVFs
forming at the same high velocity as the Ca~II DHVFs.  In particular,
blends of Ti~II HDVFs suppressed the spectrum in the blue (see
Figure~2).  The flux blocking by Ti~II HDVFs decreased with time,
which is in the right sense to cause the $B$--band light curve to rise
more quickly and fall more slowly than it otherwise would have done,
and to cause the peculiar $B - V$ evolution.  Quantitative evaluation
of the effects of line blocking by DHVFs forming in an asymmetric
matter distribution will require 3D spectrum calculations such as
those of Thomas et~al. (2003).  In any case, as an extreme case of
DHVFs SN~2000cx shows the importance of learning how much of the SN~Ia
spectroscopic and photometric diversity is caused by DHVFs.

SN~2000cx spectra contained a distinct, persistent absorption feature
near 4530~\AA\ for which the most plausible identification that we
(Thomas et~al. 2003 and Branch et~al., in preparation) can suggest is
an H$\beta$ ($\lambda$4861) DHVF, blueshifted by the same high
velocity as the Ca~II and Ti~II DHVFs.  At this blueshift the
H$\alpha$ absorption happens to fall within the red Si~II absorption.
As discussed by Thomas et~al. and Branch et~al., if H$\alpha$ and
P$\alpha$ are confined to a clump in front of the photosphere and
have source functions that are elevated relative to the
resonance--scattering source function (as they generally do in
SNe~II), then they could be difficult to see.  In principle their
spectroscopic signatures even could vanish.  The H$\beta$
identification remains tentative, however, because there is no
independent evidence for the presence of hydrogen.


\begin{figure}
\includegraphics[width=8.cm, angle=270]{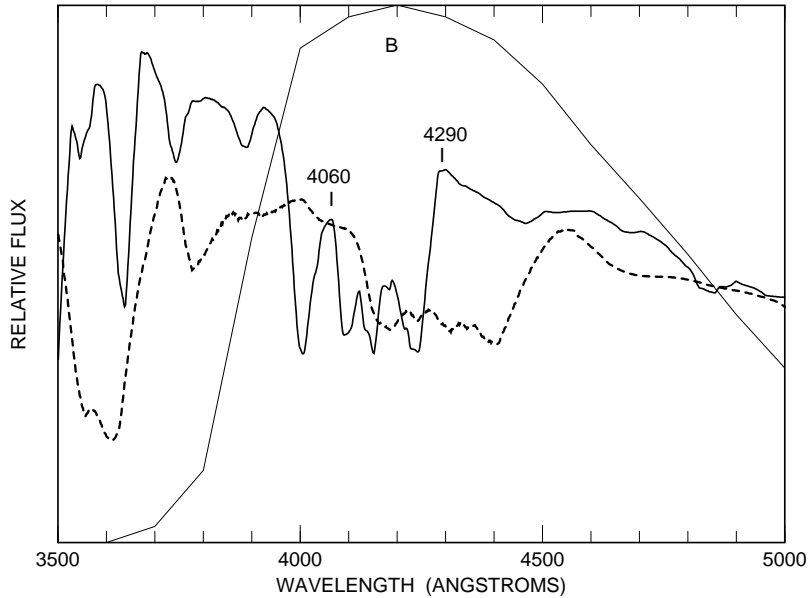}
\caption{ Two synthetic spectra having $v_{phot}=11,000$ km~s$^{-1}$
and containing only Ti~II lines are compared.  One spectrum (dashed
line) contains only undetached lines and has features like those seen
in the peculiar SN~1991bg.  The other spectrum (thick solid line)
contains only lines detached at 23,000 km~s$^{-1}$, as required to fit
the spectrum of the peculiar SN~2000cx.  The $B$--band filter function
also is shown (thin solid line). }
\end{figure}

SN~2002cx was just as surprising as SN~2000cx.  Li et~al. (2003)
observed SN~2002cx and referred to it as the ``most peculiar''
SN~Ia. Near maximum its spectra first contained features of Fe~III;
Fe~II features developed later.  Despite the fact that Si~II and S~II
features apparently were not present, Li et~al. argued that SN~2002cx
should be regarded as a Type~Ia --- the first one observed to be blue,
yet subluminous.  At maximum light the velocity at the photosphere was
only 7000 km~s$^{-1}$, making it the first good example of an LPV
({\sl low} photospheric velocity) SN~Ia.  Li et~al. called attention
to mysterious emission lines in the red part of spectra obtained three
weeks postmaximum, and to what appeared to be an unusually early
appearance of nebular emission lines two months postmaximum.  The
early development of the nebular phase was taken as an indication that
the mass ejection may have been sub--Chandraskhar.  Direct analysis
with {\bf Synow} (D. Branch et~al., in preparation) clarifies some of
these issues.  At maximum light the spectrum contained not only Fe~III
features but also weak features of Si~III, Si~II, and S~II ---
confirming the Type~Ia classification.  The spectrum also contained
Fe~II and Ca~II DHVFs forming at about 14,000 km~s$^{-1}$. The {\bf
Synow} analysis reveals that in the spectrum several weeks postmaximum
the mysterious emission lines in the red actually are blends of
P~Cygni features of permitted Fe~II and Co~II lines (see Figure~3),
and that the spectrum two months postmaximum had not gone nebular ---
it was much like the two--week postmaximum spectrum, but with stronger
Fe~II and weaker Co~II.  SN~2002cx raises several interesting
questions.  For example, if the ejected composition was dominated by
iron--peak elements, the kinetic energy per unit mass should have been
high, so why were the velocities low?  And if the mass ejection was
Chandraskhar, why was the luminosity low?  One possibility (see the
chapter by Kasen) is that SN~2002cx was a low--luminosity
SN~1991bg--like event, viewed right down the hole in the ejecta caused
by the presence of the donor star (Marietta, Burrows, \& Fryxell
2000).


\begin{figure}
\includegraphics[width=8.cm, angle=270]{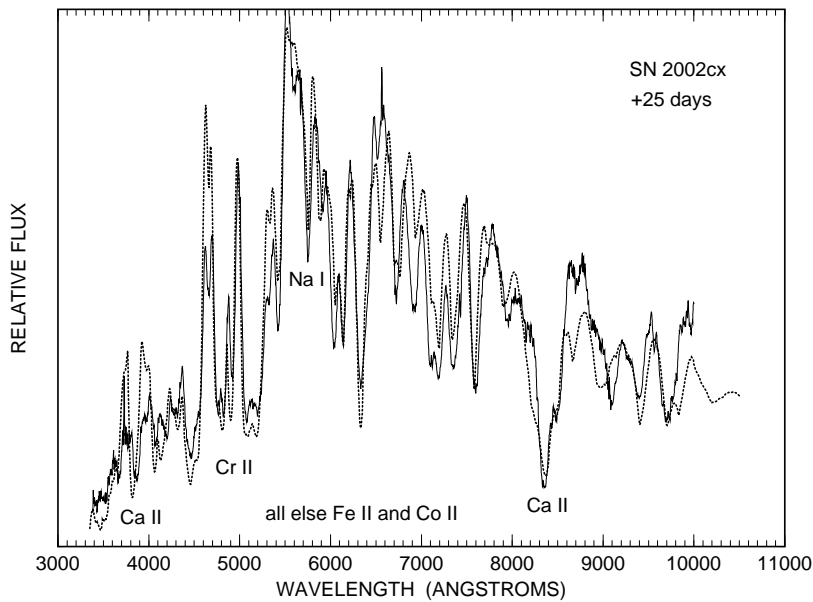}
\caption{A synthetic spectrum of having $v_{phot}=7000$ km~s$^{-1}$
  and dominated by strong permitted resonance scattering features of
  Fe~II and Co~II (dotted line) is compared with a spectrum of the
  peculiar SN~2002cx obtained by Li et~al. (2003) 25 days after
  maximum light (solid line).
}
\end{figure}

\section{Questions}

Thinking of $R(Si~II)$ as a measure of the nickel mass and of
$V_{10}(Si~II)$ as a measure of the density at high velocity may be a
useful way to think about the diversity, but clearly it is not enough.
Some of the diversity is produced by DHVFs, which we now recognize to
be not uncommon in SNe~Ia.  Weak DHVFs such as those of SN~1994D
(Hatano et~al 1999) may be caused by radial ionization variations in
smooth ejecta density distributions (H\"oflich et al. 1998; Lentz
et~al. 2001a), but what is the origin of the stronger DHVFs --- those
that appear to require density enhancements at high velocity?  The
most extreme case so far is SN~2000cx with its strong Ti~II DHVFs.
The DHVFs of SNe~2001el and 2000cx appear to be in asymmetric matter
distributions.  Are they a consequence of asymmetric matter ejection?

Has an H$\beta$ DHVF been detected in SN~2000cx?  If so, at least some
of the matter producing the DHVFs of SN~2000cx must have come from a
nondegenerate companion star.  Do any other SNe~Ia also show H$\beta$
DHVFs, at less noticeable levels?  Gerardy et~al. (2003) suggest that
strong DHVFs are produced by high--velocity density enhancements owing
to interaction with solar--abundance circumstellar matter.  If this is
correct, then it may be DHVFs, rather than narrow H$\alpha$ emission
or absorption, that are the first to reveal the presence of the
long--sought circumstellar matter associated with SN~Ia
single--degenerate progenitor systems.

Events such as SNe~2001ay, 2002cx, and 2002ic seem, so far, to be
peculiar each in their own way.  As mentioned above, the effects of
the Marietta et~al. hole in the ejecta may be responsible for some of
the peculiarities of SN~2002cx, as well as some of the other SN~Ia
diversity (see the chapter by Kasen).  SN~2002ic (Hamuy et~al. 2003;
see also the chapter by Hamuy) evidently is a very special case, since
it shows a clear signature of circumstellar hydrogen to a degree that
would not have been missed in spectra of previously observed SNe~Ia.
Note that the characteristic velocity of the H$\alpha$ emission of
SN~2002ic, $\sim1800$ km~s$^{-1}$, is much lower than the velocity of
the H$\beta$ DHVF tentatively identified in SN~2000cx ($\sim23,000$
km~s$^{-1}$).

Learning how to relate the various manifestations of observational
diversity --- the range of $R(Si~II)$ and $V_{10}(Si~II)$ values, the
DHVFs, and the individually peculiar events --- to their physical
causes --- e.g., nickel mass, flame propagation, asymmetric ejection,
asymmetric circumstellar interaction, looking down the hole --- is
looking like an increasingly challenging problem.

\bigskip\noindent {\it Acknowledgements:} I have had the benefit of
many discussions of these matters with Eddie Baron, Peter H\"oflich,
Dan Kasen, Peter Nugent, and Rollin Thomas.  It is a pleasure to take
this opportunity to thank Craig Wheeler for all he has done to make
supernova research so interesting and enjoyable.

\end{document}